\documentclass[twocolumn]{aastex631}

\usepackage{graphicx}
\usepackage[varg]{txfonts}
\usepackage{cases}
\usepackage{natbib}
\usepackage{multirow}
\usepackage{longtable}
\usepackage{graphicx}
\usepackage{booktabs}

\shorttitle{Vertical oscillations of coronal loops}
\shortauthors{Zhang et al.}

\graphicspath{{./}{figures/}}

\begin{document}

\title{Transverse vertical oscillations during the contraction and expansion of coronal loops}

\correspondingauthor{Qingmin Zhang}
\email{zhangqm@pmo.ac.cn}

\author[0000-0003-4078-2265]{Qingmin Zhang}
\affiliation{Key Laboratory of Dark Matter and Space Astronomy, Purple Mountain Observatory, CAS, Nanjing 210023, People's Republic of China}
\affiliation{Yunnan Key Laboratory of the Solar physics and Space Science, Kunming 650216, People's Republic of China}

\author{Yuhao Zhou}
\affiliation{School of Astronomy and Space Science, University of Science and Technology of China, Hefei 230026, People's Republic of China}

\author[0000-0001-7693-4908]{Chuan Li}
\affiliation{School of Astronomy and Space Science, Nanjing University, Nanjing 210023, People's Republic of China}
\affiliation{Key Laboratory of Modern Astronomy and Astrophysics (Nanjing University), Ministry of Education, Nanjing 210023, People's Republic of China}

\author[0000-0001-7540-9335]{Qiao Li}
\affiliation{School of Astronomy and Space Science, University of Science and Technology of China, Hefei 230026, People's Republic of China}

\author[0000-0002-2630-4753]{Fanxiaoyu Xia}
\affiliation{Key Laboratory of Dark Matter and Space Astronomy, Purple Mountain Observatory, CAS, Nanjing 210023, People's Republic of China}

\author[0000-0002-1190-0173]{Ye Qiu}
\affiliation{School of Astronomy and Space Science, Nanjing University, Nanjing 210023, People's Republic of China}
\affiliation{Key Laboratory of Modern Astronomy and Astrophysics (Nanjing University), Ministry of Education, Nanjing 210023, People's Republic of China}

\author[0000-0003-4787-5026]{Jun Dai}
\affiliation{Key Laboratory of Dark Matter and Space Astronomy, Purple Mountain Observatory, CAS, Nanjing 210023, People's Republic of China}

\author[0000-0003-1979-9863]{Yanjie Zhang}
\affiliation{School of Astronomy and Space Science, University of Science and Technology of China, Hefei 230026, People's Republic of China}

\begin{abstract}
In this paper, we carry out a detailed analysis of the M1.6 class eruptive flare occurring in NOAA active region 13078 on 2022 August 19.
The flare is associated with a fast coronal mass ejection (CME) propagating in the southwest direction with an apparent speed of $\sim$926 km s$^{-1}$. 
Meanwhile, a shock wave is driven by the CME at the flank. 
The eruption of CME generates an extreme-ultraviolet (EUV) wave expanding outward from the flare site with an apparent speed of $\geq$200 km s$^{-1}$.
As the EUV wave propagates eastward, it encounters and interacts with the low-lying adjacent coronal loops (ACLs), which are composed of two loops.
The compression of EUV wave results in contraction, expansion, and transverse vertical oscillations of ACLs.
The commencements of contraction are sequential from western to eastern footpoints and the contraction lasts for $\sim$15 minutes.
The speeds of contraction lie in the range of 13$-$40 km s$^{-1}$ in 171 {\AA} and 8$-$54 km s$^{-1}$ in 193 {\AA}.
A long, gradual expansion follows the contraction at lower speeds. Concurrent vertical oscillations are superposed on contraction and expansion of ACLs.
The oscillations last for 2$-$9 cycles and the amplitudes are $\leq$4 Mm. The periods are between 3 to 12 minutes with an average value of 6.7 minutes.
The results show rich dynamics of coronal loops.
\end{abstract}


\section{Introduction} \label{intro}
Coronal mass ejections (CMEs) are impulsive, large-scale eruptions of magnetic field and corona plasma into the heliosphere \citep[see][and references therein]{chen11,geo19}.
A majority of CMEs are produced by eruptions of prominences or magnetic flux ropes originating from quiet regions or active regions \citep{fan05,au10,yan18,zqm2022,zhou23}.
The speeds of CMEs have a wide range from $\sim$100 km s$^{-1}$ to $\geq$3000 km s$^{-1}$ \citep{yas04}. 
Successive stretching of the overlying magnetic field lines generates a bright, expanding front from the source region observed in extreme ultraviolet (EUV) wavelengths, 
which is named ``EIT wave" \citep{tho98,chen02,chen05,ball05}.
Meanwhile, a faster coronal Moreton wave is frequently observed to propagate ahead of the EIT wave \citep{cw11,kum13,devi22}.
It is generally accepted that an EUV wave consists of a wave-like component moving at fast magnetosonic speed \citep{will99,zhe23} and a coherent 
driven compression front related to the eruption \citep{down11}.
Occasionally, the fast-mode wave is a shock wave accompanied by a type II radio burst \citep{go09,zuc18}.

Waves and oscillations are omnipresent in the solar atmosphere \citep{and09,rud09,jess15,naka20}, such as Alfv{\'e}n waves \citep{ef07,liu19}, slow-mode \citep{wang03,xia22},
and fast-mode magnetohydrodynamics (MHD) waves \citep{er83}.
Fast-mode kink oscillations of coronal loops excited by flares were first observed by the Transition Region and Coronal Explorer (TRACE) spacecraft \citep{asch99,naka99}.
Kink oscillations could also been induced by small-scale magnetic reconnection \citep{he09},
coronal jets \citep{dai21}, prominence eruptions \citep{zim15}, and EUV waves \citep{shen12,kum13,sri13,guo15,su18,devi22}.
Coronal seismology based on kink oscillations provides an effective way of determining the 
magnetic field strength, internal Alfv{\'e}n speed, and density scale height of the oscillating loops \citep{naka01,ver04,ve08,li17,yang20,zqm20,li23}.
Interaction between EIT waves and coronal loops has been used to estimate the wave energy, which is hard to measure directly \citep{ball05,flu19}.
 
According to the polarization direction, transverse oscillations are categorized into horizontal and vertical oscillations. 
For the horizontal-polarized type, the direction of oscillation is perpendicular to the loop plane.
For the vertical-polarized type, the direction of oscillation is consistent with the loop plane \citep{ws04,wht12,ver17,zqm22a}.
\citet{mro11} reported vertical loop oscillations driven by a failed filament eruption from below.
On the other hand, \citet{ree20} detected Doppler oscillations of the magnetic tuning fork created by reconnection outflows propagating downward.
Another category of vertical oscillation is caused by implosion of large-scale, overlying coronal loops during the impulsive phase of a flare \citep{go12,sun12,sim13}.
The equilibrium positions of vertical oscillations are rapidly contracting inward at speeds of a few tens to $\geq$100 km s$^{-1}$, and there is no recovery to their original heights.
\citet{rus15} proposed a unified model (``remove-of-support" mechanism) to self-consistently explain the contraction and vertical oscillation. 
Besides, vertical loop oscillation after the strong impact of an EUV wave has been observed \citep{sri13,su18}.
\citet{mu05} carried out two-dimensional (2D) numerical simulations of vertical, kink-mode loop oscillations that are excited impulsively.
The effect of varying the initial pulse position is explored and the results are consistent with previous observations by \citet{ws04}.
Furthermore, \citet{sel10} performed 2D numerical simulations of vertical kink oscillations excited by an oscillatory driver.
Comparison with the impulsive excitation by a pressure pulse (e.g., an EUV wave) shows that attenuation of vertical kink oscillations is greatly reduced.
Energy leakage is a predominant mechanism of quick attenuation of vertical kink oscillations \citep{sel05,sel07}.
Using 3D numerical modelings, \citet{ofm15} investigated the vertical loop oscillations after the impact of a fast-mode shock wave.
The periods are very close to the observed values \citep{sri13}.

\citet{chan21} reported contraction and expansion of coronal loops induced by a nearby filament eruption.
They conjectured that a coronal wave is generated during the eruption, pushing the loops downward followed by a recovery to their initial positions.
\citet{zqm22b} studied the eruption of an EUV hot channel (flux rope) on 2022 January 20, which produced an M5.5 class flare, a fast CME, and an EUV wave.
During its propagation, the EUV wave encounters and compresses low-lying adjacent coronal loop (ACLs), resulting in rapid contraction, expansion, and vertical oscillation of ACLs. 
Inspired by the ``remove-of-support" mechanism, the authors put forward a new scenario to explain the expansion and oscillation (see their Fig. 12).
They speculated that vertical oscillation of high-lying coronal loops during contraction as a result of implosion and 
vertical oscillation of low-lying loops during expansion after the EUV wave leaves are physically the same.
Until now, vertical oscillations during both contraction and expansion motions have not been observed.

In this paper, we report multiwavelength observations of an M1.6-class flare (SOL2022-08-19T04:44) on 2022 August 19.
The eruptive flare and the associated CME initiated in NOAA active region (AR) 13078 (S27W48). 
Transverse vertical oscillations in low-lying ACLs were detected during both contraction and expansion phases. The paper is organized as follows. 
We describe observations and related data analysis in Section~\ref{data}. The results are presented in Section~\ref{res} and discussed in Section~\ref{dis}.
Finally, a brief summary is given in Section~\ref{sum}.

\begin{figure}
\includegraphics[width=0.45\textwidth,clip=]{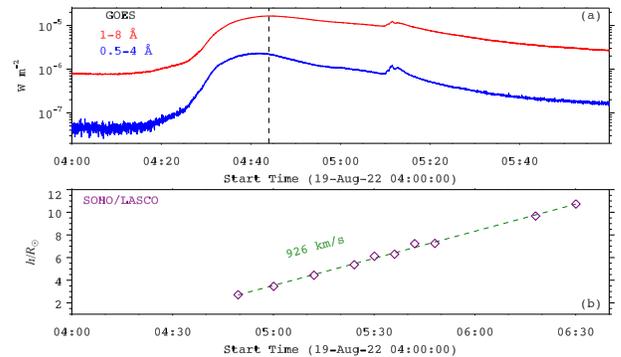}
\centering
\caption{(a) SXR light curves of the M1.6 flare in 1$-$8 {\AA} (red line) and 0.5$-$4 {\AA} (blue line). The vertical dashed line marks the peak time.
(b) Height-time plot of the CME leading edge observed by SOHO/LASCO till 06:30 UT. The apparent speed ($\sim$926 km s$^{-1}$) of the CME is labeled.}
\label{fig1}
\end{figure}

\section{Observations and data analysis} \label{data}
The M1.6 flare was observed by the Atmospheric Imaging Assembly \citep[AIA;][]{lem12} on board the Solar Dynamics Observatory (SDO) spacecraft.
AIA takes full-disk images in seven EUV (94, 131, 171, 193, 211, 304, and 335 {\AA}) and two UV (1600 and 1700 {\AA}) wavelengths 
with a spatial resolution of 1$\farcs$2 and time cadences of 12 s (EUV) or 24 s (UV).
The level 1.0 data were calibrated using \textit{aia\_prep.pro} in the \textit{Solar Software} (\textit{SSW}).
In its early phase, the flare was observed by the full-disk H$\alpha$ Imaging Spectrograph (HIS) on board the Chinese H$\alpha$ Solar Explorer \citep[CHASE;][]{li22}. 
CHASE/HIS provides H$\alpha$ spectroscopic observations with a pixel spectral resolution of 0.024 {\AA}, a spatial resolution of 1$\farcs$2, and a time cadence of one minute \citep{qiu22}.
The H$\alpha$ images were coaligned with the AIA 304 {\AA} images with an accuracy of $\sim$0$\farcs$5.
Soft x-ray (SXR) fluxes of the flare in 0.5$-$4 {\AA} and 1$-$8 {\AA} were detected by the GOES-16 spacecraft with a cadence of 1 s.
The CME was observed by the Large Angle and Spectrometric Coronagraph \citep[LASCO;][]{bru95} 
on board the SOHO spacecraft \footnote{cdaw.gsfc.nasa.gov/CME\_list/UNIVERSAL/2022\_08/univ2022\_08.html}.
The radio dynamic spectra during eruption was obtained from the Australia-ASSA ground-based station\footnote{www.e-callisto.org} with a cadence of 0.25 s and a frequency coverage of 17$-$87 MHz.

Figure~\ref{fig1}(a) shows the SXR light curves of the flare in 1$-$8 {\AA} (red line) and 0.5$-$4 {\AA} (blue line), respectively.
The SXR emission starts to rise at $\sim$04:14 UT, peaks at $\sim$04:44 UT (black dashed line), and declines gradually until $\sim$05:50 UT. Therefore, the lifetime of the flare is over 1.5 hr.

\begin{figure*}
\includegraphics[width=0.90\textwidth,clip=]{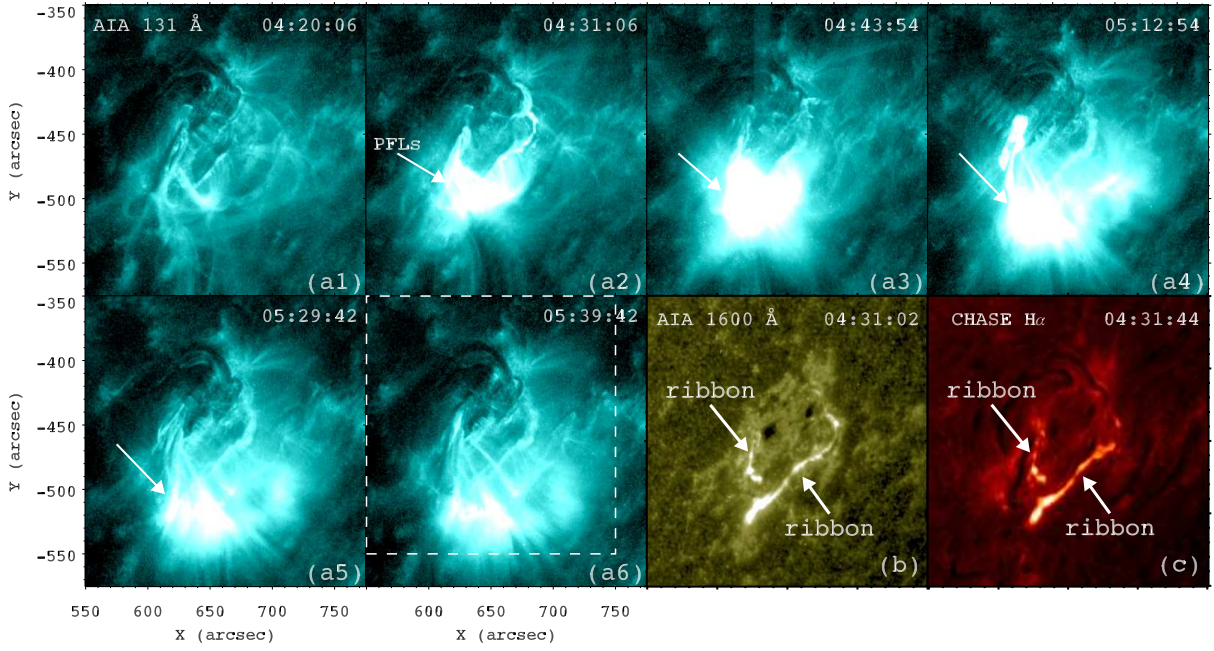}
\centering
\caption{(a1-a6): AIA 131 {\AA} images to illustrate the evolution of the flare. The white arrows point to the hot post-flare loops (PFLs). 
In panel (a6), the white dashed box denotes the FOV of panels (b-c).
(b-c): Flare ribbons in the chromosphere observed by AIA 1600 {\AA} and CHASE H$\alpha$ line center, respectively.
An animation showing the PFLs in AIA 131 {\AA} is available.
It covers a duration of 80 minutes from 04:20 UT to 05:40 UT on 2022 August 19. The entire movie runs for $\sim$4 s.
(An animation of this figure is available.)}
\label{fig2}
\end{figure*}

Figure~\ref{fig2}(a1-a6) show six AIA 131 {\AA} images to illustrate the evolution of the flare (see also the online animation).
The arrows point to the bright and hot post-flare loops (PFLs). A magnetic flux rope or hot channel was not detected before and during the eruption \citep{rg11,zqm22b}.
Figure~\ref{fig2}(b-c) show the bright flare ribbons in the chromosphere observed in 1600 {\AA} and H$\alpha$ line center (6562.8 {\AA}) at 04:31 UT, respectively.
The western ribbon is much longer and brighter than the eastern one.

\begin{figure*}
\includegraphics[width=0.90\textwidth,clip=]{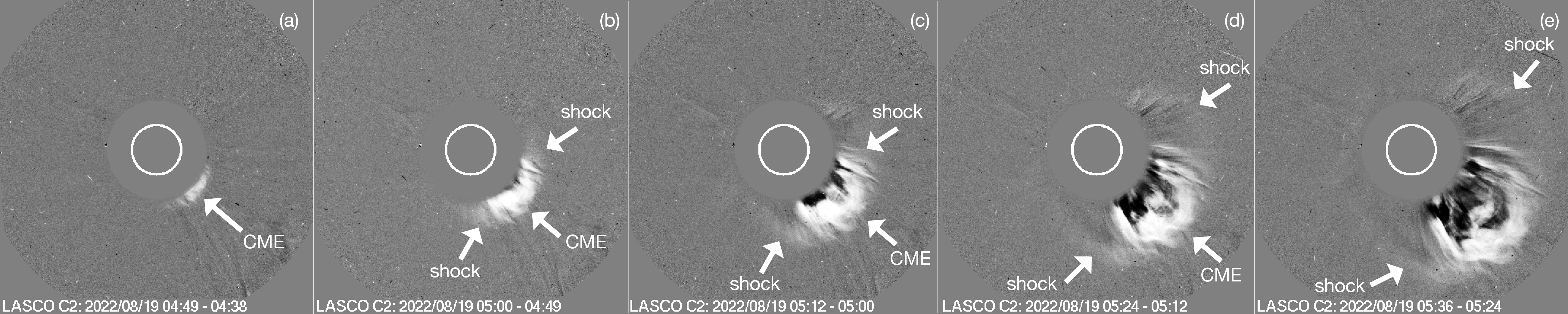}
\centering
\caption{Running-difference images of the CME observed by LASCO/C2 during 04:49$-$05:36 UT.
The white arrows point to the CME and CME-driven shock.}
\label{fig3}
\end{figure*}

Figure~\ref{fig3} shows white-light (WL) images of the CME and shock wave observed by LASCO/C2 coronagraph.
The CME first appears at 04:49:30 UT and propagates in the southwest direction with a central position angle of $\sim$225$\degr$. 
The angular width reaches $\sim$104$\degr$ due to the CME-driven shock.
The height evolution of CME during 04:49$-$06:30 UT is plotted with purple diamonds in Figure~\ref{fig1}(b). 
A linear fitting of the height as a function of time results in an apparent speed of $\sim$926 km s$^{-1}$.
Taking the projection effect into account, the true speed of CME reaches $\sim$1246 km s$^{-1}$, which is sufficient to drive a shock.
 
\begin{figure*}
\includegraphics[width=0.90\textwidth,clip=]{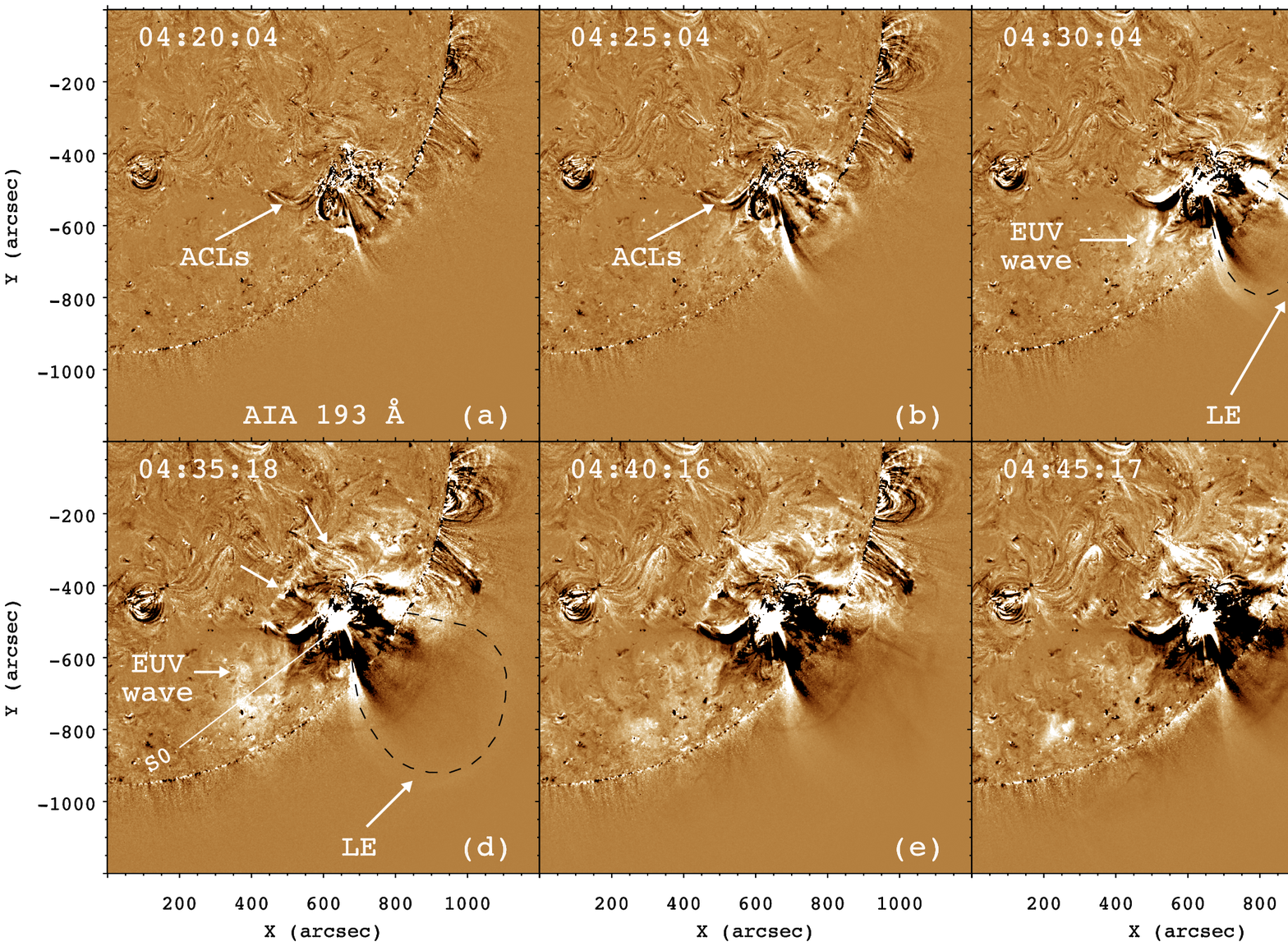}
\centering
\caption{AIA 193 {\AA} base-difference images to illustrate the formation and propagation of an EUV wave.
In panels (a-b), the arrows point to the adjacent coronal loops (ACLs). In panels (c-d), the arrows point to the EUV wave and CME leading edge (LE).
An artificial slice (S0) with a length of 371$\arcsec$ is used to investigate the evolution of EUV wave.
An animation showing the EUV wave in AIA 193 {\AA} is available. It covers a duration of 25 minutes from 04:20 UT to 04:45 UT on 2022 August 19. The entire movie runs for $\sim$1 s.
(An animation of this figure is available.)}
\label{fig4}
\end{figure*}

\begin{figure}
\includegraphics[width=0.45\textwidth,clip=]{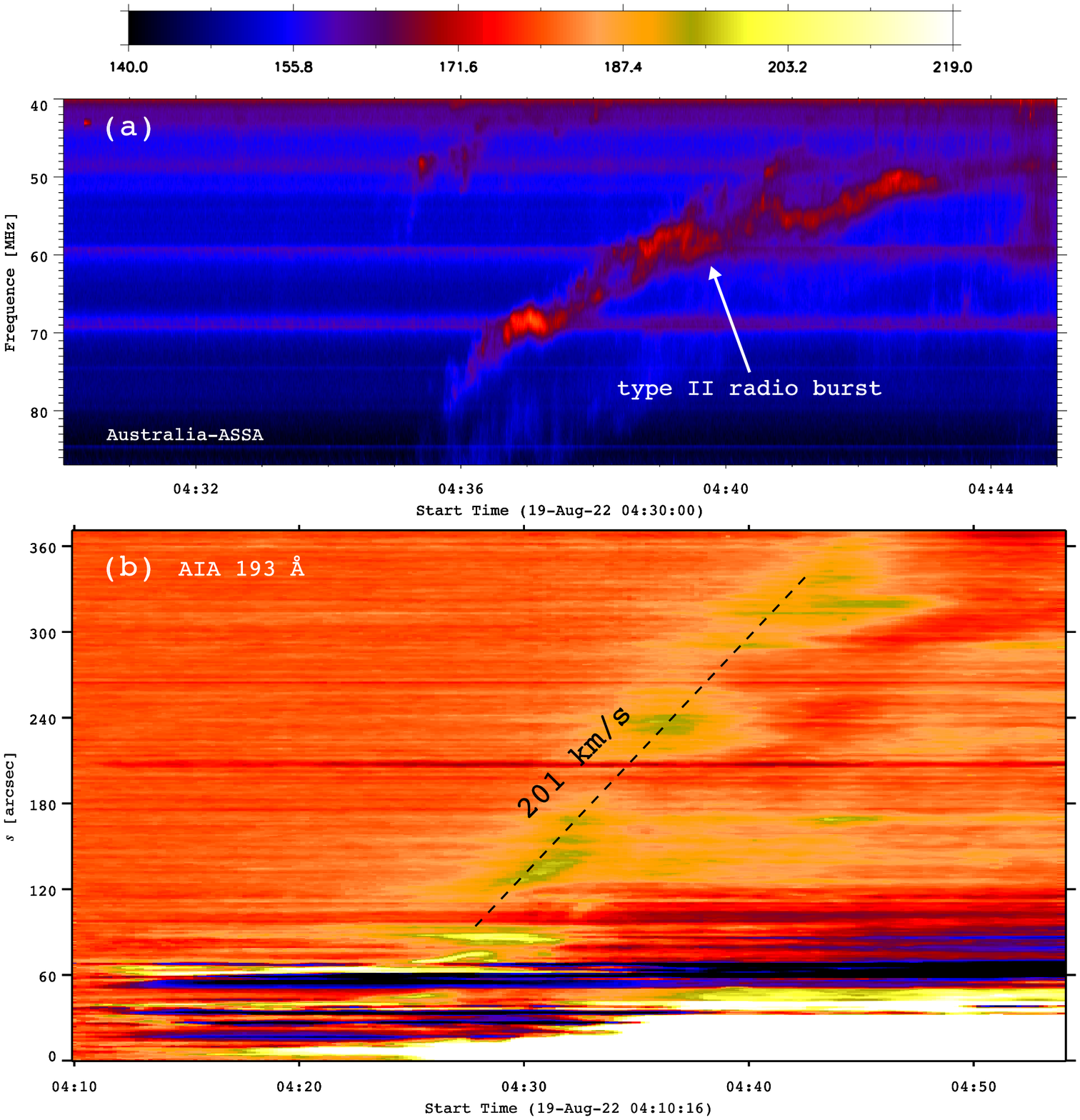}
\centering
\caption{(a) Radio dynamic spectra of the eruptive flare observed by the Australia-ASSA station. The arrow points to the type II radio burst related to the CME-driven shock.
(b) Time-distance plot of S0 in 193 {\AA}. $s=0$ and $s=371\arcsec$ denote the northwest and southeast endpoints of S0, respectively.
The apparent speed ($\sim$201 km s$^{-1}$) of the EUV wave is labeled.}
\label{fig5}
\end{figure}

Figure~\ref{fig4} shows six AIA 193 {\AA} base-difference images during 04:20$-$04:45 UT (see also the online animation). 
The arrows indicate the leading edge (LE) of the CME in its early phase.
As the CME rises and propagates in the southwest direction, a bright EUV wave expands outward from the flare site and detaches from the CME bubble, leaving behind a dark dimming region.
To study the evolution of EUV wave, a straight slice (S0) with a length of 371$\arcsec$ is selected in panel (d).
The time-distance plot of S0 is displayed in Figure~\ref{fig5}(b). The EUV wave front propagates from $\sim$04:26 UT to $\sim$04:45 UT at an apparent speed of $\sim$201 km s$^{-1}$ on the disk.

Figure~\ref{fig5}(a) shows the radio dynamic spectra of the eruptive flare observed by the Australia-ASSA station, featuring a type II radio burst related to the shock.
The frequency drifts slowly from $\sim$80 to $\sim$45 MHz during 04:34:30$-$04:43:30 UT \citep{zuc18}.
Therefore, the shock is formed at the early phase of CME evolution below 2$R_\sun$, which is probably due to its initial overexpansion \citep{pat10}.
Coincidence of the EUV wave and type II radio burst suggests that the bright EUV wave front is the imprint of the CME-driven shock on the surface.

\section{Results} \label{res}
\begin{figure}
\includegraphics[width=0.45\textwidth,clip=]{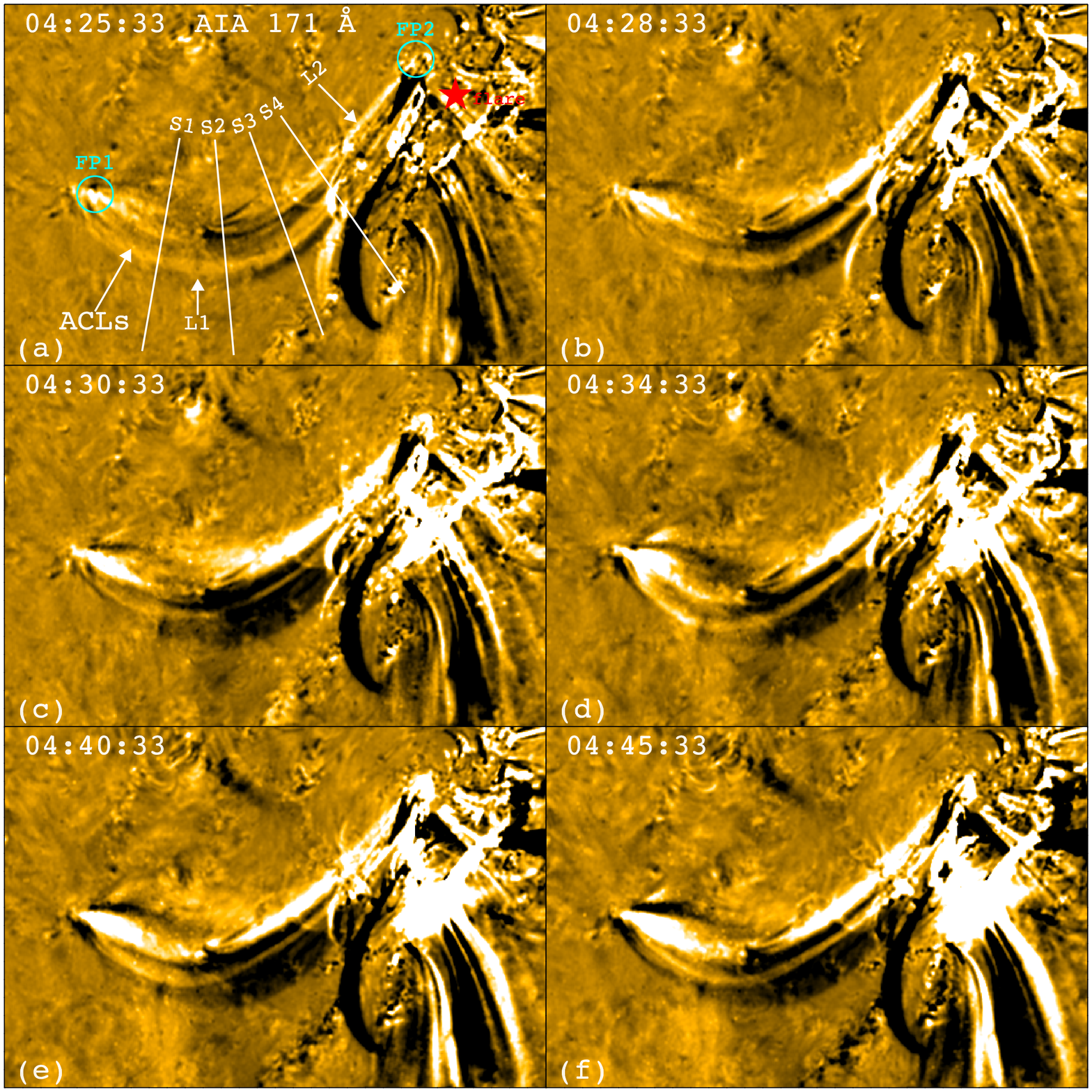}
\centering
\caption{Snapshots of the AIA 171 {\AA} images during the contraction, expansion, and oscillation phases.
In panel (a), a red star marks the flare site, and two cyan circles mark the footpoints of ACLs, i.e., FP1 and FP2. The ACLs are composed of two loops, L1 and L2.
Four slices (S1, S2, S3, S4) with the same length of 87 Mm are used to investigate the evolution of ACLs.
An animation showing the passage of an EUV wave and the subsequent expansion and transverse vertical oscillation of ACLs in AIA 171 {\AA} is available.
It covers a duration of 35 minutes from 04:25 UT to 05:00 UT on 2022 August 19. The entire movie runs for $\sim$5 s.
(An animation of this figure is available.)}
\label{fig6}
\end{figure}

\begin{figure}
\includegraphics[width=0.45\textwidth,clip=]{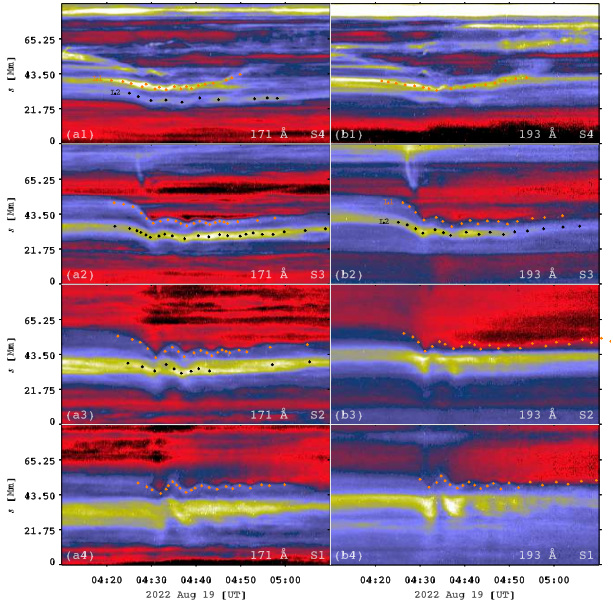}
\centering
\caption{Time-distance plots of S1-S4 in AIA 171 {\AA} (left panels) and 193 {\AA} (right panels).
$s=0$ and $s=87$ Mm signify the north and south endpoints of the slices. The manually extracted trajectories of L1 and L2 are denoted with orange and black ``+'' symbols, respectively.}
\label{fig7}
\end{figure}

In Figure~\ref{fig4}(a-b), the base-difference images show the ACLs next to the flare region. As the EUV wave expands outward, it sweeps the ACLs.
In Figure~\ref{fig6}, six AIA 171 {\AA} base-difference images illustrate the interaction between the EUV wave and ACLs (see also the online animation).
It is clear that as the EUV wave arrives from the flare site and compresses the ACLs, the interaction causes contraction, expansion, and oscillation of ACLs \citep{zqm22b}.
In panel (a), the footpoints (FP1 and FP2) of ACLs are marked by two cyan circles, which are $\sim$193$\arcsec$ apart. 
The ACLs are composed of two loops, L1 and L2, with L1 being higher and longer than L2.
Four slices (S1-S4) with the same length of 87 Mm are selected to investigate the evolution of ACLs. 
Time-distance plots of the four slices in 171 and 193 {\AA} are displayed in the left and right panels of Figure~\ref{fig7}, respectively.
The manually extracted trajectories of L1 and L2 are denoted with orange and black ``+'' symbols, respectively. The contractions of ACLs commence after 04:20 UT and persist until $\sim$04:36 UT. 
It is worth noting that the start times of contraction are sequential from S4 to S1 since S4 is closer to the flare site.
The speeds of contraction of L1 are larger than those of L2, which is most probably due to that L1 is higher than L2 and responds to the compression of EUV wave earlier than L2.
Afterwards, the ACLs expand gradually until 05:15 UT and the speeds of expansion are evidently lower than those of contraction.
The final heights of ACLs are equal to or lower than the initial heights. Hence, the whole process lasts for $\sim$55 minutes.
Interestingly, vertical oscillations are superposed on the contraction and expansion, 
which is different from previously reported situations of oscillation during contraction \citep{go12,sun12,sim13} as well as oscillation during expansion \citep{zqm22b}.
The oscillation of L1 is obvious along all slices in 171 and 193 {\AA}.
On the contrary, the oscillation of L2 is distinguishable along S2-S4 in 171 {\AA} (panels (a1-a3)), while it is clear only along S3 in 193 {\AA} (panel (b2)).

\begin{deluxetable*}{ccc|ccc|ccc|c|ccc}
\tablecaption{Fitted parameters of the trend in Equation~(\ref{eqn-1}) and the related periods ($P$) and damping times ($\tau_d$) of vertical oscillations. \label{tab-1}} 
\tablecolumns{13}
\tablenum{1}
\tablewidth{0pt}
\tablehead{
\colhead{$\lambda$} &
\colhead{Slice} &
\colhead{Loop} &
\colhead{$c_{1}$} &
\colhead{$t_{1}$} &
\colhead{$a_{1}$} &
\colhead{$c_{2}$} &
\colhead{$t_{2}$} &
\colhead{$a_{2}$} &
\colhead{$d$} &
\colhead{$P$} &
\colhead{$\tau_d$} &
\colhead{$\tau_d/P$} \\
\colhead{{(\AA})} &
\colhead{} &
\colhead{} &
\colhead{(Mm)} &
\colhead{(s)} &
\colhead{(s)} &
\colhead{(Mm)} &
\colhead{(s)} &
\colhead{(s)} &
\colhead{(Mm)} &
\colhead{(min)} &
\colhead{(min)} &
\colhead{}
}
\startdata
171 & S1 & L1 & 1.73 & 115.67 &  2.04 &  61.82 & 4713.04 & 1438.41 & 66.41                & 6.5 & 8.3 & 1.3 \\
171 & S2 & L1 & 5.06 &  338.44 &  99.14 &    148.61 &  5367.20 &  1450.24 &    156.08      & 6.4 & 10.8 & 1.7 \\
171 & S2 & L2 &  8.50 &    -81.03 &    732.88 &     78.50 &   7595.76 &   4389.96 &     80.63     & 7.5 & 10.8 & 1.4 \\
171 & S3 & L1 &  6.06 &    357.18 &    120.68 &     32.48 &   2959.28 &    524.60 &     40.45     & 7.2 & 14.2 & 2.0 \\
171 & S3 & L2 &  3.25 &    314.77 &    130.43 &    565.58 &   7436.12 &   1728.17 &    569.72     & 6.5 & 25.7 & 3.9 \\       
171 & S4 & L1 &  4.04 &    345.48 &    352.94 &     54.70 &   2813.15 &    988.94 &     57.17    & 3.0, 6.0 & 25.7 & 8.6, 4.3 \\
171 & S4 & L2 &  2.30 &    120.46 &     25.81 &      3.42 &   2576.69 &   1169.69 &      6.71    & 12.0 & 9.7 & 0.8 \\
\hline
193 & S1 & L1 & 0.30 &    169.00 &      3.88 &      1.60 &   2155.64 &    315.52 &      4.83      & 5.8 & 26.3 & 4.5 \\
193 & S2 & L1 &  3.89 &    160.87 &     60.14 &      2.27 &   2231.14 &    464.54 &      9.31  & 6.4 & 34.7 & 5.4 \\ 
193 & S3 & L1 & 8.28 &    121.69 &    208.45 &      2.02 &   1704.12 &    191.20 &     12.81 & 6.5 & 10.8 & 1.7 \\
193 & S3 & L2 & 3.57 &    186.61 &    132.26 &      2.13 &   1866.29 &    286.71  &     7.07 & 6.5 & 20.8 & 3.2 \\
193 & S4 & L1 & 6.25 &    428.07 &    476.08 &      9.01 &   1327.76 &    973.07 &      9.13   & ... & ...  & ... \\
\enddata
\end{deluxetable*}

\begin{figure}
\includegraphics[width=0.45\textwidth,clip=]{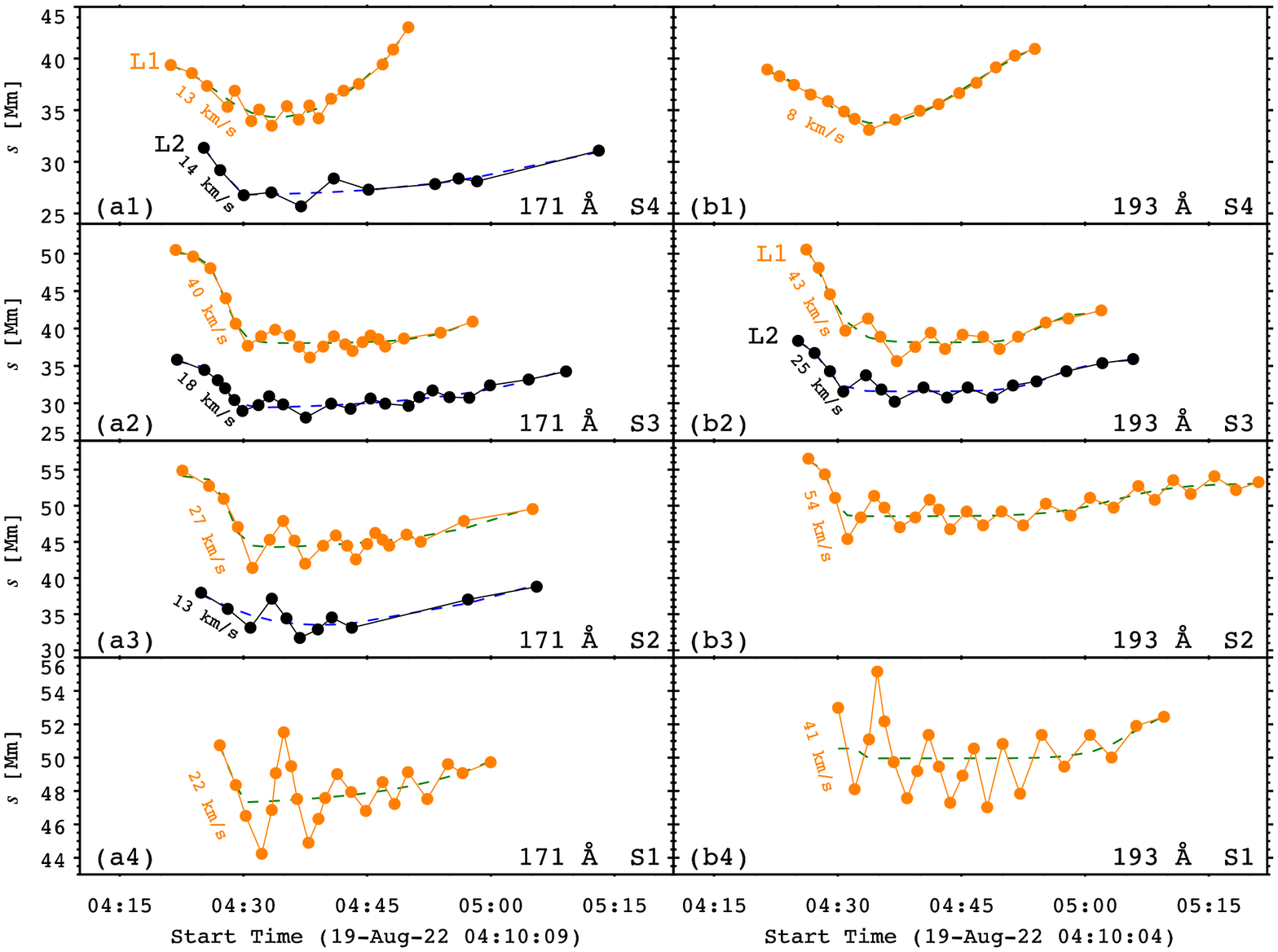}
\centering
\caption{Trajectories of ACLs (orange circles for L1 and black circles for L2) along S1-S4 in 171 {\AA} (left panels) and 193 {\AA} (right panels). 
The green and blue dash lines represent the fitted trends using Equation~(\ref{eqn-1}).
The speeds of initial contraction are labeled.}
\label{fig8}
\end{figure}

The trajectories of L1 and L2 along S1-S4 are plotted with orange and black circles in Figure~\ref{fig8}.
To precisely derive the background trend including the contraction and expansion, we apply the following function of $t$:
\begin{equation} \label{eqn-1}
  h(t)=-c_{1}\tanh(\frac{t-t_{1}}{a_{1}}) + c_{2}\tanh(\frac{t-t_{2}}{a_{2}}) + d,
\end{equation}
where $c_{1}$, $t_{1}$, $a_{1}$, $c_{2}$, $t_{2}$, $a_{2}$, and $d$ are seven free parameters. 
The first and second terms represent contraction \citep{rus15} and expansion \citep{zqm22b}, respectively. $t_{1}$ and $t_{2}$ denote times after the onsets of contraction.
The curve fittings are performed by using the \textit{mpfit.pro} program in \textit{SSW}.
The fitted parameters are listed in Table~\ref{tab-1} and the background trends are superposed with green and blue dashed lines in Figure~\ref{fig8}.
It is clear that the trends are smooth and the fittings are satisfactory in most cases.

\begin{figure*}
\includegraphics[width=0.90\textwidth,clip=]{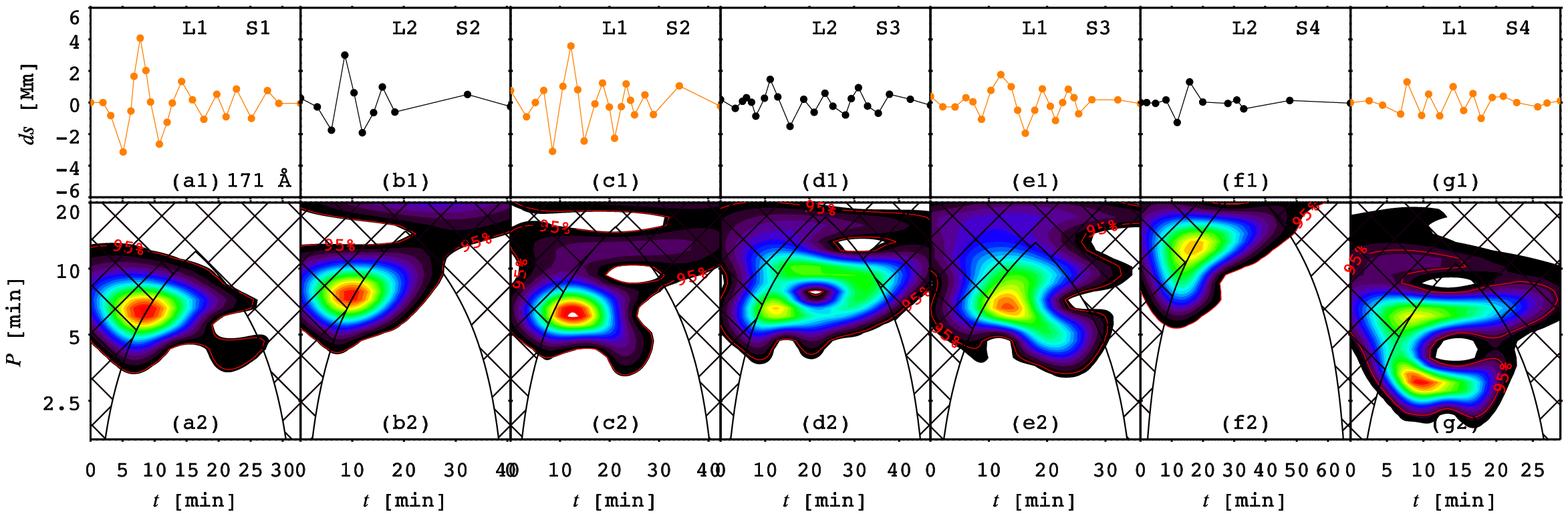}
\centering
\caption{Morlet wavelet transforms (bottom panels) of the associated detrended trajectories in 171 {\AA} (top panels).
The horizontal axes represent the times after onsets of contraction.
The red lines represent the 95\% confidence level. The black lines show the cone of influence.}
\label{fig9}
\end{figure*}

In Figure~\ref{fig8}, the associated detrended trajectories of ACLs are obtained after subtracting the background trends. 
In Figure~\ref{fig9}, the top panels show the detrended trajectories of L1 and L2 in 171 {\AA}. It is obvious that vertical oscillations of the loops last for 2$-$9 cycles and the amplitudes are $\leq$4 Mm.
Morlet wavelet transforms of the detrended trajectories are displayed in the bottom panels of Figure~\ref{fig9}.
The corresponding periods of oscillations are between 3 and 12 minutes, which are listed in the eleventh column of Table~\ref{tab-1}.
The multiple periods suggest possible existence of harmonics in loop oscillations \citep{de07,vand07,duck18}.
To derive the damping times of the vertical oscillations, the detrended trajectories are fitted with the function \citep{naka99}:
\begin{equation} \label{eqn-2}
  y(t)=A_{0}\sin(\frac{2\pi}{P}(t-t_{0})+\phi_{0})\exp(-\frac{t-t_{0}}{\tau_{d}}),
\end{equation}
where $A_{0}$ and $\phi_{0}$ denote the initial amplitude and phase at $t_{0}$, $P$ is the period derived from the wavelet transform, and $\tau_{d}$ represents the damping time.
The fitted values of $\tau_{d}$ and ratios of $\tau_{d}/P$ are listed in the last two columns of Table~\ref{tab-1}.

\begin{figure}
\includegraphics[width=0.45\textwidth,clip=]{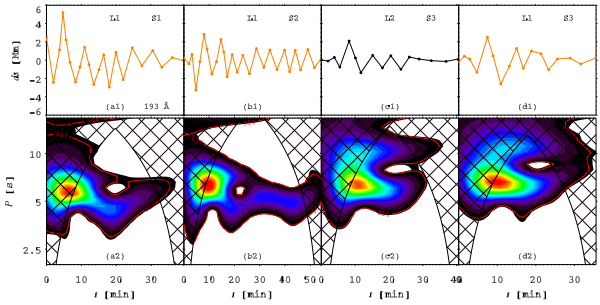}
\centering
\caption{Morlet wavelet transforms (bottom panels) of the associated detrended trajectories in 193 {\AA} (top panels).
The horizontal axes represent the times after onsets of contraction.}
\label{fig10}
\end{figure}

The top panels of Figure~\ref{fig10} show detrended trajectories of L1 and L2 in 193 {\AA}.
Morlet wavelet transforms of the detrended trajectories are displayed in the bottom panels of Figure~\ref{fig10}.
The corresponding periods of oscillations are between 5.8 and 6.5 minutes.
It is noted that the oscillation along S4 in 193 {\AA} (Figure~\ref{fig8}(b1)) is marginal. Hence, wavelet transform was not carried out.
Likewise, the corresponding $\tau_{d}$ and $\tau_{d}/P$ in 193 {\AA} are listed in the last two columns of Table~\ref{tab-1}.
The damping times of all ACLs in 171 and 193 {\AA} range from 8 to 35 minutes with a median value of $\sim$14 minutes.
The ratio of $\tau_{d}/P$ range from 0.8 to 8.5 with a median value of $\sim$2.0, which is close to that of horizontal oscillations \citep{zqm20,dai21}.
The average and standard deviation of all periods of ACLs are 6.7 and 2 minutes, respectively.
This average period is close to the periods of transverse loop oscillation on 2001 April 15 \citep{ver04} and 2001 June 15 \citep{asch02}.
Assuming a semi-circular shape of the ACLs, the average loop length ($L$) is estimated to be $\sim$220 Mm, which is also equal to the loop length on 2001 April 15 \citep{ver04}.
The phase speed of vertical oscillation is calculated to be $C_k=2L/P\approx1094$ km s$^{-1}$, which is very close to the value reported by \citet{naka01}.
Assuming a density ratio of $\sim$0.1 between the external and internal plasma \citep{naka99,chen15}, the average internal Alfv\'{e}n speed of the loops is estimated to be $\sim$811 km s$^{-1}$.

\begin{deluxetable*}{c|c|c|c|c|c|c|c|c}
\tablecaption{Comparison of the three events, including the flare class, association with a shock, initial contraction speeds in 304, 171, and 193 {\AA}, 
period of loop oscillations, and ratio of $\tau_{cie}/\tau_{osc}$, respectively. \label{tab-2}} 
\tablecolumns{9}
\tablenum{2}
\tablewidth{0pt}
\tablehead{
\colhead{Date} &
\colhead{Flare} &
\colhead{Shock} &
\colhead{$V_{304}$} &
\colhead{$V_{171}$} &
\colhead{$V_{193}$} &
\colhead{$P$} &
\colhead{$\tau_{cie}/\tau_{osc}$} &
\colhead{Ref.} \\
\colhead{} &
\colhead{} &
\colhead{} &
\colhead{(km s$^{-1}$)} &
\colhead{(km s$^{-1}$)} &
\colhead{(km s$^{-1}$)} &
\colhead{(min)} &
\colhead{} &
\colhead{}
}
\startdata
2022/01/20 & M5.5 & Yes & 44 & 71 & 71 & 4.4$\pm$0.2 & $\ll$1 & \citet{zqm22b} \\
2022/08/19 & M1.6 & Yes & ... & 13$-$40 & 8$-$54 & $\sim$6.7 & $\sim$1 & this study \\
2013/03/16 & ... & No & ... & 2$-$10 & 3$-$11 & ... & $\gg$1 & \citet{chan21} \\
\enddata
\end{deluxetable*}

\section{Discussion} \label{dis}
To explain the contraction and vertical oscillation of the overlying coronal loops as a result of coronal implosion, \citet{rus15} employs a novel equation:
\begin{equation} \label{eqn-3}
  \frac{d^2x}{dt^2}+\omega^2(x-x_0(t))+2\omega\kappa\frac{dx}{dt}=0,
\end{equation}
where $x(t)$ denotes the displacement of the loop, $x_0(t)$ denotes the equilibrium position as a function of $t$, $\omega$ denotes the frequency of the undamped oscillation,
and $\kappa$ is the damping ratio. 
The response of overlying loop depends on the ratio of change-in-equilibrium time scale ($\tau_{cie}$) to the period of oscillation ($\tau_{osc}$), i.e., $\tau_{cie}/\tau_{osc}$.
The loop impulsively reaches a new equilibrium position with oscillation when $\tau_{cie}\ll \tau_{osc}$.
In contrast, the loop collapses slowly without oscillation when $\tau_{cie}\gg \tau_{osc}$.
Moderate collapse with oscillation takes place when the two parameters are comparable (see their Fig. 4).
The speed of contraction ($V_{con}$) is the highest when $\tau_{cie}\ll \tau_{osc}$ and lowest when $\tau_{cie}\gg \tau_{osc}$. In other words, $\tau_{cie}$ is inversely proportional to $V_{con}$.
Consequently, the vertical oscillation seems to be stealth during the impulsive collapse and becomes discernible on the condition of moderate collapse.
\citet{zqm22b} studied the contraction, expansion, and vertical oscillation of ACLs close to AR 12929 on 2022 January 20. The final heights of ACLs exceed their initial heights before flare.
They concluded that Equation~(\ref{eqn-3}) can interpret not only the contraction and oscillation of overlying coronal loops, but also the expansion and oscillation of ACLs.
Moreover, an innovative cartoon is proposed to illustrate the whole process (see their Fig. 12).
In the current study, the values of $V_{con}$ in 171 {\AA} are between 13 and 40 km s$^{-1}$ with a median value of 18 km s$^{-1}$ (see Figure~\ref{fig8} and Table~\ref{tab-2}).
The values of $V_{con}$ in 193 {\AA} are between 8 and 54 km s$^{-1}$ with a median value of 41 km s$^{-1}$, which are considerably lower than the value (71 km s$^{-1}$) on 2022 January 20.
That is to say, the values of $\tau_{cie}$ in the current event are much longer than that on 2022 January 20, 
which may explain the coherent vertical oscillations during both contraction and expansion phases of ACLs.

Using the multiwavelength observations from SDO/AIA, \citet{chan21} investigated the filament eruption and evolutions of two sets of adjacent loop systems on 2013 March 16.
The filament eruption was related to a partial halo CME and a weak coronal wave, but was not associated with a detectable x-ray flare.
Both sets of coronal loops underwent in-phase contraction first and then expand roughly back to their original positions without kink oscillation.
The speeds of contraction vary from 2 to 10 km s$^{-1}$ in 171 {\AA} and from 3 to 11 km s$^{-1}$ in 193 {\AA} (see Table~\ref{tab-2}).
Therefore, the speeds of contraction are the lowest and the value of $\tau_{cie}$ is the longest on 2013 March 16, which might interpret the absence of oscillation during the contraction.
In brief, the three events are representative of three cases of coronal loop dynamics in response to the impact of an EUV wave in vertical direction with different conditions, 
including rapid contraction followed by expansion and oscillation, coherent oscillation superposed on contraction and expansion, pure contraction and expansion without oscillation.
These works convincingly reveal the rich dynamics of coronal loops.

\section{Summary} \label{sum}
In this paper, we perform a detailed analysis of the M1.6 class eruptive flare occurring in AR 13078 on 2022 August 19.
The flare is associated with a fast CME propagating in the southwest direction with an apparent speed of $\sim$926 km s$^{-1}$. Meanwhile, a shock wave is driven by the CME at the flank.
The early evolution of CME is accompanied by a type II radio burst, suggesting the shock wave is formed during the impulsive acceleration of CME.
The eruption of CME generates an EUV wave expanding outward from the flare site with an apparent speed of $\geq$200 km s$^{-1}$.
As the EUV wave propagates eastward, it encounters and interacts with the low-lying ACLs, which are composed of two loops (L1 and L2).
The compression of EUV wave results in contraction, expansion, and transverse vertical oscillations of ACLs.
The start times of contraction are sequential from the western to eastern footpoints and the contraction lasts for $\sim$15 minutes.
The speeds of contraction lie in the range of 13$-$40 km s$^{-1}$ in 171 {\AA} and 8$-$54 km s$^{-1}$ in 193 {\AA}. A long, gradual expansion follows the contraction at lower speeds.
Concurrent vertical oscillations are superposed on contraction and expansion of ACLs, which has rarely been reported before.
The oscillations last for 2$-$9 cycles and the amplitudes are $\leq$4 Mm. The periods are between 3 to 12 minutes with an average value of 6.7 minutes.
According to the loop length and periods, the average kink speed ($\sim$1094 km s$^{-1}$) and internal Alfv\'{e}n speed ($\sim$811 km s$^{-1}$) of the ACLs are estimated.

In the future, in-depth investigations of the interaction between EUV waves and coronal loops are highly desirable, especially using high-resolution observations from 
the Extreme Ultraviolet Imager \citep[EUI;][]{ro20} on board Solar Orbiter \citep[SO;][]{mu20}.
State-of-the-art MHD numerical simulations are greatly expected to explain the loop dynamics \citep{down21,wang21,guo23}.

\begin{acknowledgments}
The authors appreciate the referee for valuable suggestions and comments.
SDO is a mission of NASA\rq{}s Living With a Star Program. AIA data are courtesy of the NASA/SDO science teams.
The CHASE mission is supported by China National Space Administration. The e-Callisto data are courtesy of the Institute for Data Science FHNW Brugg/Windisch, Switzerland.
This work is supported by the National Key R\&D Program of China 2021YFA1600500 (2021YFA1600502) 
and Yunnan Key Laboratory of Solar Physics and Space Science under the number YNSPCC202206.
J.D. is supported by the Special Research Assistant Project CAS.
\end{acknowledgments}


\end{document}